\documentclass[%
 aip,
 amsmath,amssymb,
 reprint,%
]{revtex4-1}

\usepackage{graphicx}
\usepackage{dcolumn}
\usepackage{bm}

\usepackage[utf8]{inputenc}
\usepackage[T1]{fontenc}
\usepackage{mathptmx}
\usepackage{etoolbox}
\usepackage{xcolor}

\makeatletter
\def\@email#1#2{%
 \endgroup
 \patchcmd{\titleblock@produce}
  {\frontmatter@RRAPformat}
  {\frontmatter@RRAPformat{\produce@RRAP{*#1\href{mailto:#2}{#2}}}\frontmatter@RRAPformat}
  {}{}
}%
\makeatother
\begin{document}

\preprint{AIP/123-QED}

\title[Sample title]{ Thin channel Ga\textsubscript{2}O\textsubscript{3} MOSFET with 55 GHz f\textsubscript{MAX} and >100 V breakdown}
\author{Chinmoy Nath Saha}
\author{Abhishek Vaidya}
\author{Noor Jahan Nipu}
 \affiliation{ Electrical Engineering, University at Buffalo, Buffalo, New York 14240, USA}
\author{Lingyu Meng},\author{Dong Su Yu},
\affiliation{ 
Electrical and Computer Engineering, Ohio State University, Columbus, OH 43210, USA}
\author{Hongping Zhao}
\affiliation{ 
Electrical and Computer Engineering, Ohio State University, Columbus, OH 43210, USA} 

\author{Uttam Singisetti}
 \affiliation{ Electrical Engineering, University at Buffalo, Buffalo, New York 14240, USA}
 
 \email{uttamsin@buffalo.edu}

\date{\today}

\begin{abstract}
This letter reports a highly scaled 90 nm gate length $\mathrm{\beta} $-Ga\textsubscript{2}O\textsubscript{3} (Ga\textsubscript{2}O\textsubscript{3}) T-gate MOSFET with  power gain cut off frequency (f\textsubscript{MAX}) of 55 GHz. The 60 nm thin epitaxial Ga\textsubscript{2}O\textsubscript{3} channel layer was grown by molecular beam epitaxy (MBE) while the highly doped (n++) source/drain regions were regrown using metal organic chemical vapour deposition (MOCVD). Maximum on current (I\textsubscript{DS, MAX}) of 160 mA/mm and trans-conductance (g\textsubscript{m}) around 36 mS/mm was measured at  V\textsubscript{DS} = 10 V for L\textsubscript{SD} = 1.5 $\mathrm{\mu }$m device. Transconductance and on current are limited by high channel sheet resistance (R\textsubscript{sheet}). Gate-drain breakdown voltage of 125 V was measured for L\textsubscript{GD} = 1.2 $\mathrm{\mu }$m. We extracted 27 GHz current gain cut-off frequency (f\textsubscript{T}) and 55 GHz f\textsubscript{MAX} for 20 V drain bias for unpassivated devices. The reported f\textsubscript{MAX} is the highest for Ga\textsubscript{2}O\textsubscript{3}. While no current collapse was seen initially for both drain and gate lag measurements for 500 ns pulse, moderate current collapse was observed after DC, RF measurements caused by electrical stressing. No large signal RF data was extracted due to a lack of proper tuning of the input (high S11) in the load-pull setup. However, after repeated DC and large signal measurement trials, we found that both f\textsubscript{T} and f\textsubscript{MAX} degraded significantly which was correlated to high-frequency g\textsubscript{m} collapse. Despite this, we calculated a high f\textsubscript{T}. V\textsubscript{BR} product of 3.375 THz.V  which is comparable with state-of-art GaN HEMTs. This figure of merit suggests that Ga\textsubscript{2}O\textsubscript{3} could be a potential candidate for X-band application.
\end{abstract}

\maketitle

$\mathrm{\beta} $-Ga\textsubscript{2}O\textsubscript{3} (Ga\textsubscript{2}O\textsubscript{3}) is an ultrawide bandgap semiconductor  with favorable materials properties\cite{green2022beta} for next-generation power and RF applications. The predicted breakdown field (8 MV/cm)\cite{green2022beta,green20163}  and calculated saturation velocity\cite{ghosh2017ab} supports the candidacy of Ga\textsubscript{2}O\textsubscript{3} for high-frequency switching and high power RF amplifier applications. $\mathrm{\beta} $-Ga\textsubscript{2}O\textsubscript{3} MOSFETs with multi-kV breakdown voltages have been reported \cite{kezeng2018,tetzner2019lateral,arkka1,zhang2022ultra,farzana2021vertical}, while heterostructure FET (HFET), and diodes with average breakdown field strength of 5.5 MV/cm\cite{kalarickal2021beta,sahabeta2} has been reported demonstrating the maturity of the technology.
Modulation doped $\mathrm{\beta}$-(Al\textsubscript{x}Ga\textsubscript{1-x})\textsubscript{2}O\textsubscript{3}/Ga\textsubscript{2}O\textsubscript{3} HFET \cite{zhang2018demonstration} and highly scaled (< 200 nm) MOSFETs\cite{chabak2018sub} and MESFETs \cite{xia2019} have been demonstrated to showcase the high-frequency performance. In our previous works, we reported f\textsubscript{T} = 30 GHz in  AlGaO/GaO HFETs\cite{vaidya2021enhancement,saha2022temperature} and  f\textsubscript{MAX}= 48 GHz using scaled T gate MOSFET\cite{sahabeta2}. Large signal RF performance has been published for L- band  \cite{moser2020pulsed}. However, f\textsubscript{MAX} > 50 GHz is necessary for S and X band applications. 

It is necessary to reduce the parasitic source resistance to achieve higher frequencies. The contact regrowth process has been reported in gallium oxide FETs\cite{vaidya2021enhancement,saha2022temperature, sahabeta2} as an effective way to to reduce contact resistance. In our previous report, we found high regrowth interface resistance\cite{sahabeta2} limiting device performance by increasing source resistance. Careful surface treatment could reduce this regrowth interface resistance problem. 
Traps in the gate and gate-drain access region can limit device RF performance by introducing current collapse known as DC-RF dispersion \cite{sahabeta2}.  Ex-situ passivation can eliminate the access region traps but traps under the gate are unaffected by passivation \cite{vaidya2021temperature}.

In this letter, we report highly scaled 90 nm T-gate Ga\textsubscript{2}O\textsubscript{3} MOSFET with improved MOCVD contact regrowth process to reduce the interface resistance. Pre-cleaning of the wafer before low-temperature n++ regrowth gave lower interface resistacne. We used atomic layer deposited Al\textsubscript{2}O\textsubscript{3} as the gate dielectric, no current collapse was observed in unpassivated devices. As a result, a peak f\textsubscript{T} of 27 GHz and f\textsubscript{MAX} = 55 GHz is obtained in these devices. With a gate-to-drain breakdown voltage of 120 V, the device shows f\textsubscript{MAX} = 55 GHz and breakdown voltage (V\textsubscript{BR}) > 100 V which has been reported only in a few state-of-art AlGaN/GaN HFETs \cite{zhang2018high,ranjan2014high,medjdoub2013record}.

\begin{figure}
\centerline{\includegraphics[width=1.1\columnwidth]{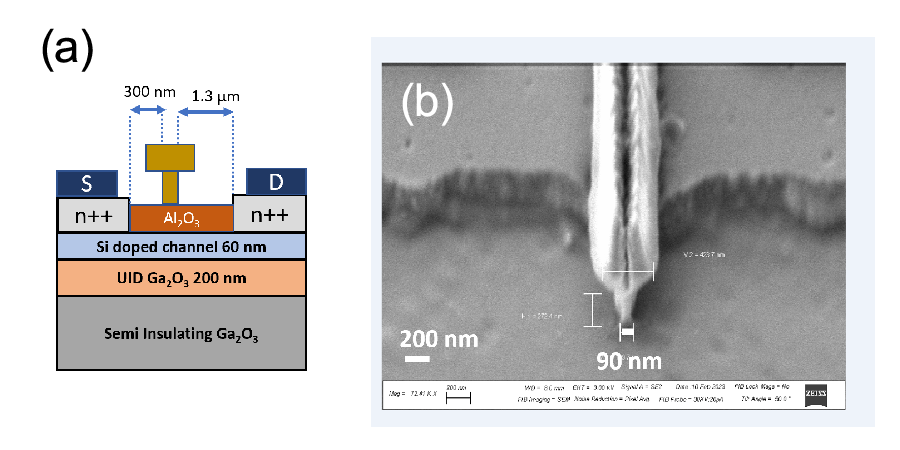}}
\vspace*{-2mm}
\caption{(a) Cross section schematic of a MOSFET (b) Magnified SEM image of a fabricated device showing 90 nm T gate with 450 nm Gate hat}
\label{fig1mos}
\end{figure}

The device layers and structure is shown in Fig. \ref{fig1mos} (a). The growth details and device fabrication process is described in detail in previous reports \cite{sahabeta2, vaidya2019structural}. Before regrowth, the sample was submerged in 1:3 HCl:DI water solution for 15 minutes to remove any atmospheric contaminants \cite{arkka1}.  Next, 80 nm highly Si doped (1 X 10\textsuperscript{20} cm\textsuperscript{-3}) n++ layer was grown  using MOCVD at a lower 650 \textsuperscript{0}C. Unlike the previous report\cite{sahabeta2}, a 20 nm aluminum oxide gate dielectric was used in these MOSFETs. SEM image of the device is shown in  Fig. \ref{fig1mos} (b), a T-gate was used to improve the gate resistance.

From the capacitance-voltage characteristics (Supplementary material), a sheet carrier charge density (2.9 
x 10\textsuperscript{12} cm\textsuperscript{-2}) was extracted. Transfer length method (TLM) on n++ regrowth layer gave a 0.045 $\mathrm{\Omega}$ mm lateral contact resistance (R\textsubscript{c,n++}) between metal and n++ layer and  a  sheet resistance (R\textsubscript{sheet,n++}) of 181 $\mathrm{\Omega}$/$\mathrm{\square}$.  TLM structure on  n++ layer through the channel was also fabricated to calculate total contact (R\textsubscript{C}) resistance to the channel. We extracted a low metal to channel contact resistance (R\textsubscript{C}) of   0.624 $\mathrm{\Omega}$ mm and a channel sheet resistance (R\textsubscript{sheet, ch}) of 28 K$\mathrm{\Omega}$/$\mathrm{\square}$.  Lower contact resistance (R\textsubscript{C}) compared to previous report\cite{sahabeta2} suggests that the surface treatment and low temperature regrowth were helpful. \cite{bhattacharyya2021multi,arkka1}

\begin{figure}
\centerline{\includegraphics[width=1.1\columnwidth]{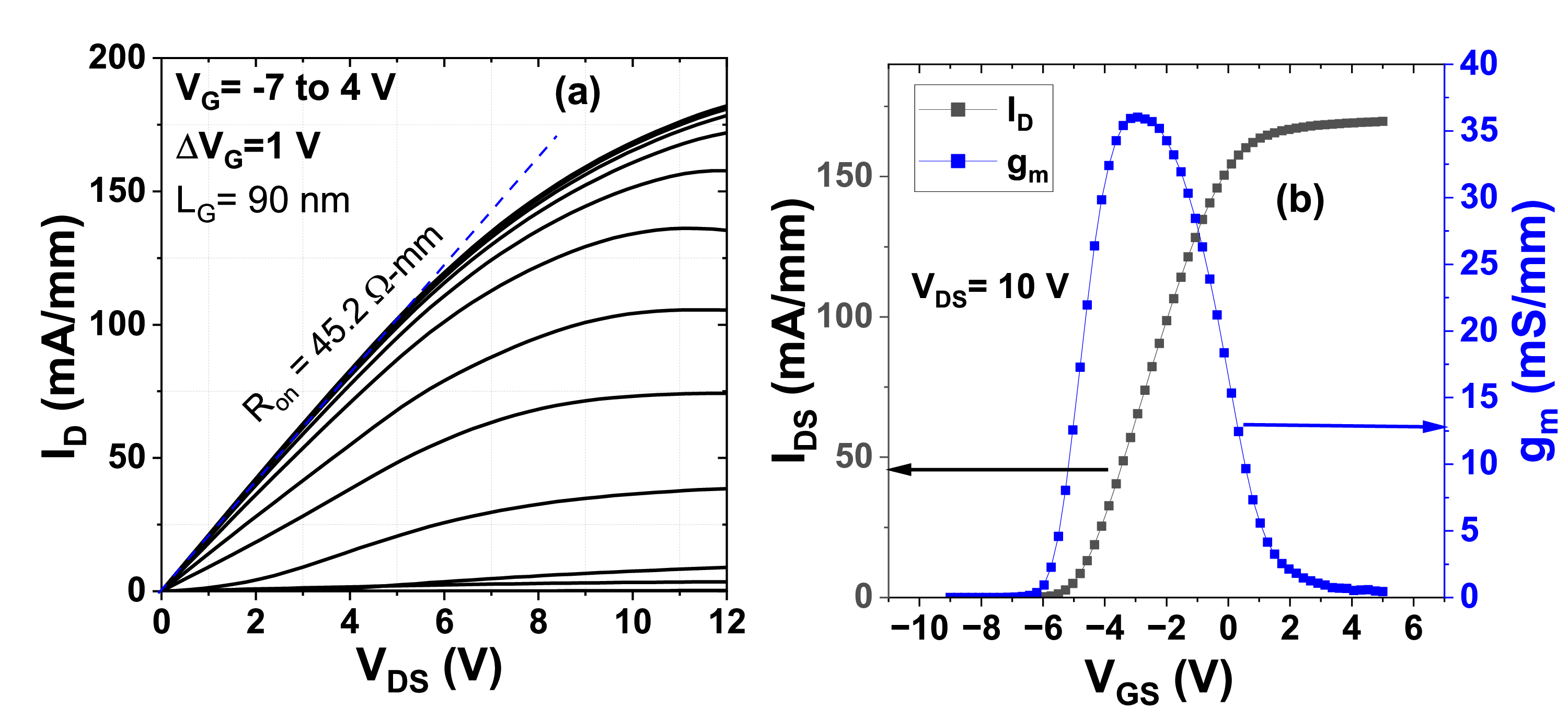}}
\vspace*{-2mm}
\caption{(a) I\textsubscript{D}- V\textsubscript{DS} output curve for a L\textsubscript{G}= 90 nm device (b)I\textsubscript{D}- V\textsubscript{GS} transfer curve at V\textsubscript{DS}= 10 V showing 36 mS/mm transconductance  }
\label{fig2mos}
\end{figure}
DC I\textsubscript{D}-V\textsubscript{DS} output curves shows peak I\textsubscript{DS, MAX} = 182 mA/mm with 45.2 $\mathrm{\Omega}$ mm at V\textsubscript{DS} = 12 V (Fig. \ref{fig2mos} (a)). Peak transconductance (g\textsubscript{m}) was found 37 mS/mm at 10 V drain bias (Fig. \ref{fig2mos} (b)). However, only 10 V drain bias was used for transfer curve in order to avoid stressing the device. Higher R\textsubscript{on} and lower g\textsubscript{m} for L\textsubscript{G} < 100 nm can be attributed to higher channel sheet resistance which increases source resistance (R\textsubscript{S}) compared to the previous report. 
\begin{figure}
\centerline{\includegraphics[width=1.1\columnwidth]{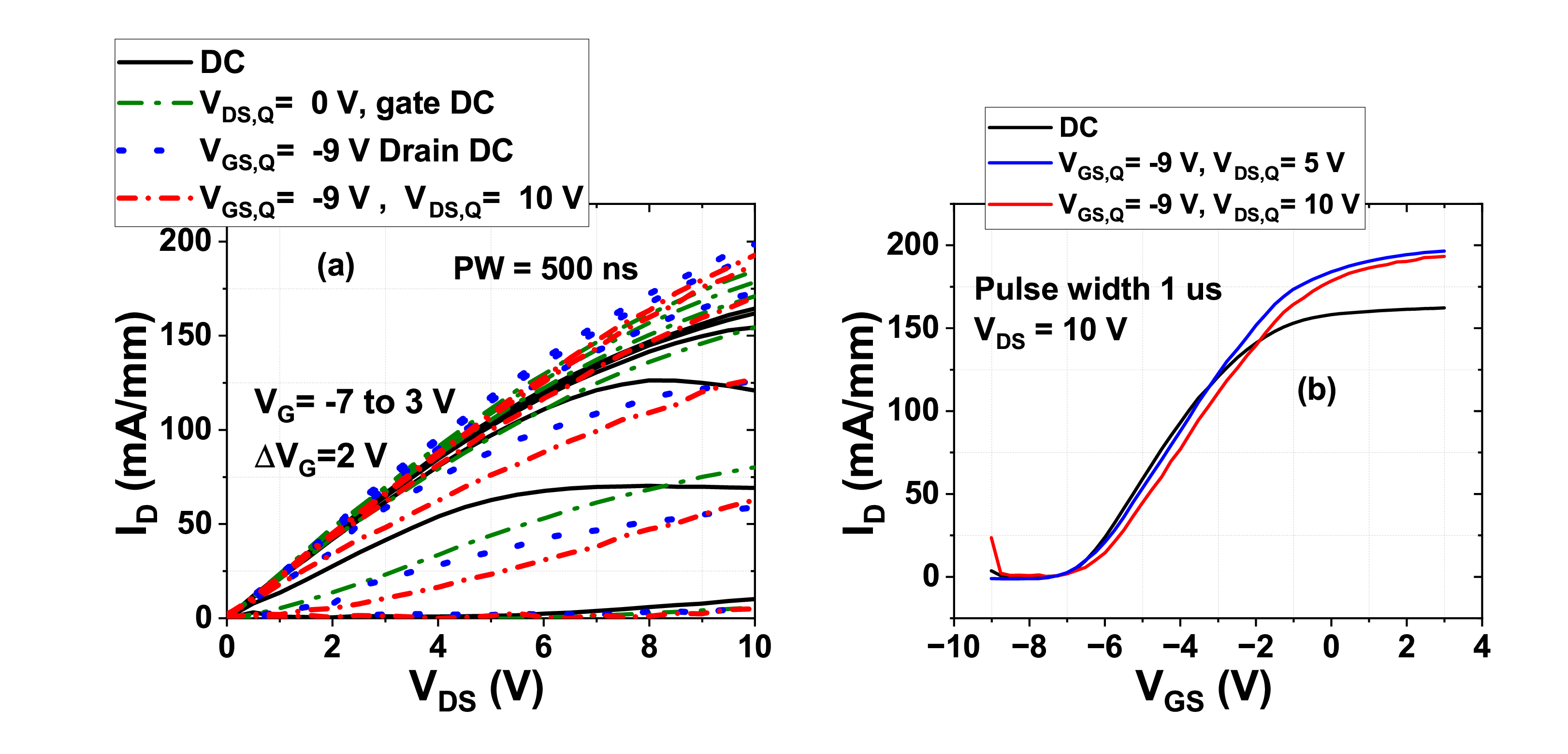}}
\vspace*{-2mm}
\caption{(a) I\textsubscript{D}- V\textsubscript{DS} output curve for different gate and drain quiescent bias points showing no current collpase (b) I\textsubscript{D}- V\textsubscript{GS} transfer curve of the device showing no current collapse at higher gate-drain quiescent point for 1 $\mu$ s pulse width}
\label{fig3mos}
\end{figure}
Three terminal off-state breakdown measurement was carried out in  air using B1505A power device analyzer. The device was biased at V\textsubscript{GS} = -15 V which is below V\textsubscript{TH}.  We observed catastrophic breakdown at V\textsubscript{DS} = 110 V (supplementary material), which corresponds to gate-to-drain breakdown voltage (V\textsubscript{BR}) of 125 V (V\textsubscript{DS}- V\textsubscript{GS}). It results in 1.18  MV/cm average breakdown field (E\textsubscript{AVG}) for 1.26 $\mathrm{\mu}$m L\textsubscript{GD} spacing. Lower E\textsubscript{AVG} can be explained due to air breakdown and lack of external passivation and field management.

Pulsed-IV measurements were carried out using Auriga-AU5 pulse voltage system with low duty cycle to reduce self-heating. No current collapse was observed and pulsed current is higher than DC at high V\textsubscript{DS} (Fig. \ref{fig3mos} (a)) for both gate and drain lag measurements. The increased current can be attributed to reduced self heating effect.  Absence of current collapse means that there is possibly no traps under gate or in gate-drain access region \cite{vaidya2021temperature,saha2022temperature}. Pulsed I\textsubscript{D}-V\textsubscript{GS} transfer curves shows no shift in threshold voltage for high V\textsubscript{DG,Q} quiescent bias points (Fig. \ref{fig3mos} (b)). This suggests that there are possibly no traps under gate or in the access region in the as deposited Al\textsubscript{2}O\textsubscript{3}. However, after repeated DC, RF and large signal measurements, we observed current collapse under pulsed measurements (See Supplementary Material). Electrical stressing may have introduced traps either in the oxide/channel or oxide/air interface. An extrinsic passivation may be necessary to reduce this DC-RF dispersion caused by electrical stressing the device.

\begin{figure}
\centerline{\includegraphics[width=0.8\columnwidth]{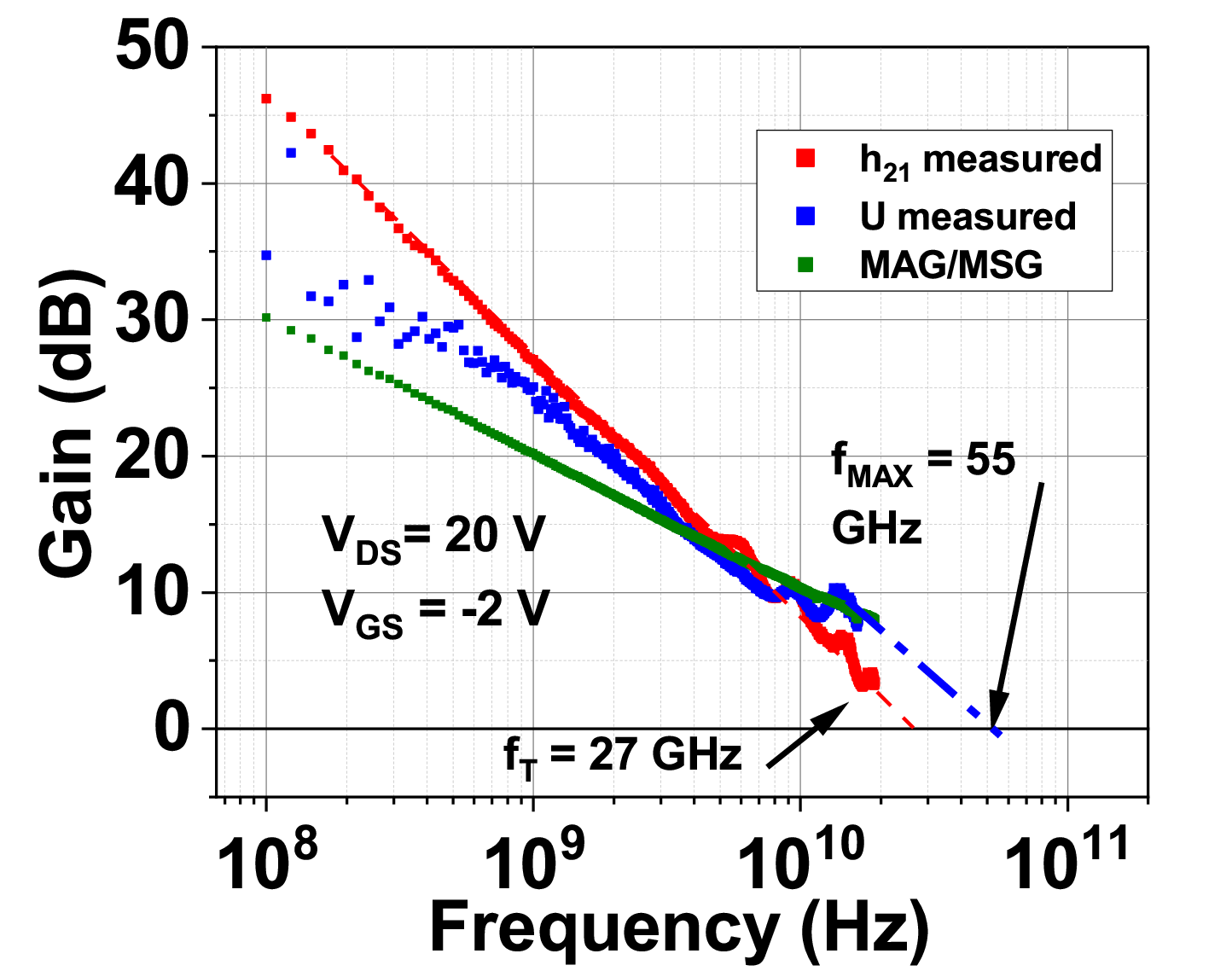}}
\vspace*{-2mm}
\caption{Measured small signal performance of a test device (L\textsubscript{G}= 90 nm) showing f\textsubscript{T} = 27 GHz and f\textsubscript{MAX}= 55 GHz}
\label{fig4mos}
\end{figure}

Small-signal analysis was performed from 100 MHz to 19 GHz using Keysight ENA 5071C Vector Network Analyzer (VNA). A sapphire calibration standard was used to calibrate The VNA by SOLT technique. An isolated open-pad device structure on the same wafer was utilized to de-embed parasitic pad capacitance \cite{koleeandembed}. Short circuit current gain (h\textsubscript{21}), Mason's unilateral gain (U) and MAG/MSG have been plotted at V\textsubscript{DS} = 12 V and V\textsubscript{GS} = 2 V bias points for L\textsubscript{G} = 90 nm device. After extrapolating to 0 dB, we found current gain cut-off frequency (f\textsubscript{t}) of $\sim$~ 27 GHz. The power gain cut off frequency (f\textsubscript{MAX}) is $\sim$~ 55 GHz (Fig. \ref{fig4mos}) which extrapolated from 20dB/decade slope from the Masons' unilateral gain (U). 
f\textsubscript{MAX} value reported here is the highest among Ga\textsubscript{2}O\textsubscript{3} FETs. We calculated the expected intrinsic f\textsubscript{T} using the geometrically calculated gate source capacitance C\textsubscript{GS} and measured DC g\textsubscript{m} (See supplemnatary materials). We assumed half of the channel thickness (30 nm) for C\textsubscript{GS} calculation. We also calculated extrinsic f\textsubscript{T} by considering R\textsubscript{S}, R\textsubscript{D} calculated from channel sheet resistance and contact resistance. These calculations \cite{tasker1989importance} gave an intrinsic and extrinsic f\textsubscript{T} of 29 GHz and 25.5 GHz respectively. This calculation gives further credence to the measured f\textsubscript{T} and f\textsubscript{MAX}. In our previous report \cite{sahabeta2}, we found DC-RF dispersion and reduced high frequency g\textsubscript{m} was the primary cause of reduced f\textsubscript{T}. 

\begin{figure}
\centerline{\includegraphics[width=1.1\columnwidth]{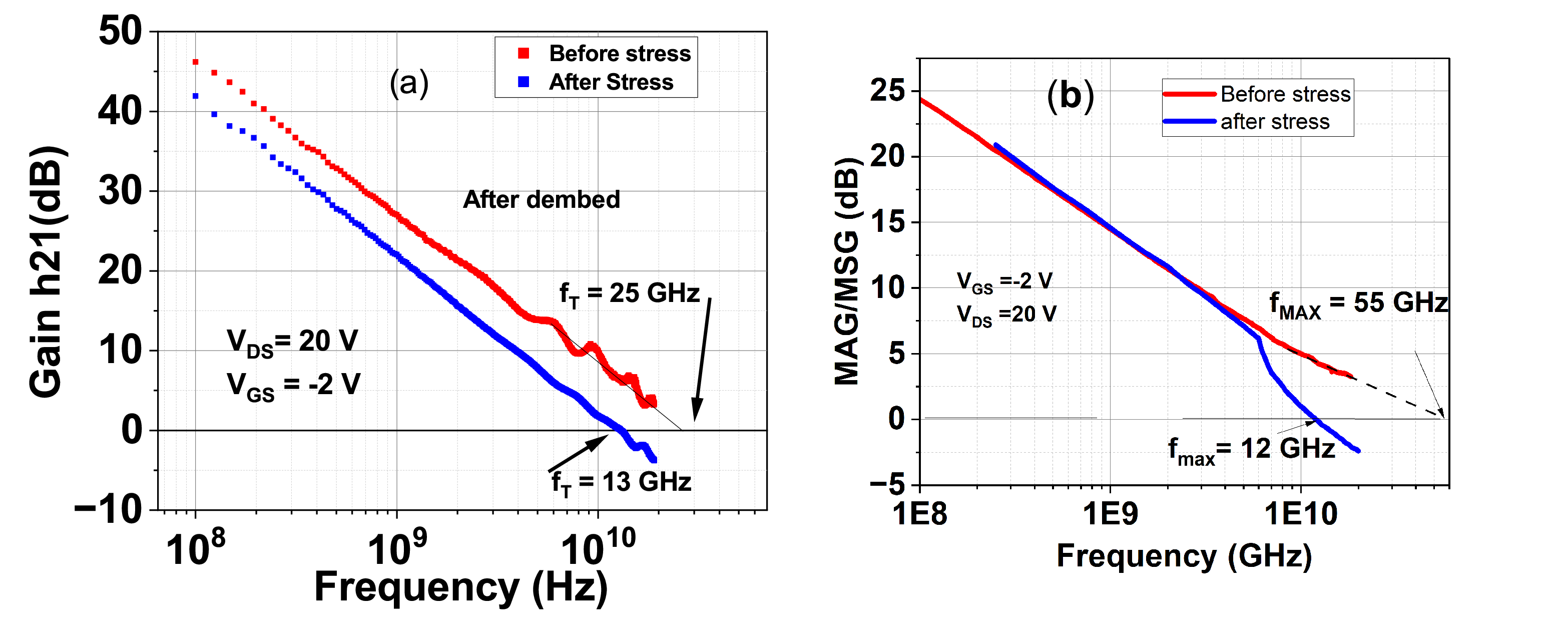}}
\vspace*{-2mm}
\caption{Degradation of  (a) f\textsubscript{T} and (b) f\textsubscript{MAX} after DC, large signal RF measurement (stress)}
\label{fig5mos}
\end{figure}
\begin{figure}
\centerline{\includegraphics[width=0.9\columnwidth]{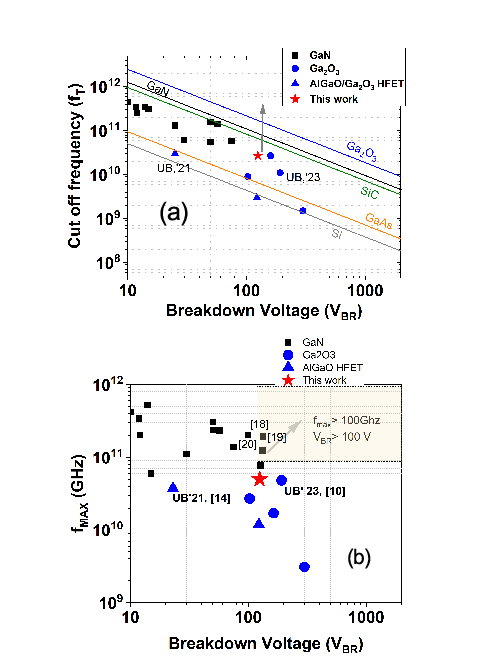}}
\vspace*{-2mm}
\caption{(a) f\textsubscript{T} vs V\textsubscript{Br} benchmark plot with GaN and other $\mathrm{\beta} $-Ga\textsubscript{2}O\textsubscript{3} devices, (b) f\textsubscript{MAX} vs V\textsubscript{Br} benchmark plot with GaN and other $\mathrm{\beta} $-Ga\textsubscript{2}O\textsubscript{3} devices,  }
\label{fig6mos}
\end{figure}
We calculated f\textsubscript{MAX} based on T gate dimension\cite{tasker1989importance}. Although we do not have any test structure to calculate exact gate resistance (R\textsubscript{G}), nonetheless our calculation results in f\textsubscript{MAX}  ~ 90 GHz which is higher than our measured data. The discrepancy can be attributed to the higher gate resistance arising from thin metal films and errors in unilateral gain (U) extraction from measured s-parameters. f\textsubscript{MAX}/f\textsubscript{T} ratio of 1.8 has been extracted which is similar to our previous report\cite{sahabeta2} and other reports (3 to 5) \cite{zhang2018demonstration,kamimura2020delay,green2017beta} in the literature.

Large signal measurements were attempted on the MOSFETs, however good input matching could not be achieved, thus no large signal data is available. Moreover, we found degradation of both f\textsubscript{T} and f\textsubscript{MAX} after repeated DC and large-signal RF measurements. f\textsubscript{T} dropped to 13 GHz from 25 GHz and f\textsubscript{MAX} dropped to 12 GHz after DC and large signal measurements (Fig. \ref{fig5mos}(a) and (b)). Similar characteristics was also observed for all devices; repeated measurements degraded device RF performance significantly (See Supplementary materials). After electrical stressing, the RF g\textsubscript{m} was lower than DC g\textsubscript{m} resulting in the f\textsubscript{T} and f\textsubscript{MAX} degradation. Pulsed I-V measurements showed current collapse after stressing. Possible trap introduction after repeated DC and large signal measurement may be the primary reason for this. It is noted that these devices did not have the typical external passivation that is used in RF GaN HEMTs. Successful external passivation could mitigate this degradation by electrical stressing. A more careful analysis is necessary to really understand the origin of the degradation of the device. 

Nevertheless, we calculated Johnson's Figure of Merit (JFOM) based on the breakdown voltage and f\textsubscript{T} (un-stressed values). With 125 V breakdown voltage (V\textsubscript{BR}) and 27 GHz f\textsubscript{T}, a  f\textsubscript{T}.V\textsubscript{BR} product of 3.375 GHz.V is achieved, which is comparable to the state-of-art GaN HEMTs (Fig.\ref{fig6mos}(a)). We also bench-marked f\textsubscript{MAX} and breakdown voltage (V\textsubscript{BR}) of our device with GaN HEMTs.As seen in Fig. \ref{fig6mos}(b), this is the only $\mathrm{\beta} $-Ga\textsubscript{2}O\textsubscript{3} device that shows f\textsubscript{MAX} 55 GHz and V\textsubscript{BR} > 100 V (Fig. \ref{fig6mos}(b)) except for a few AlGaN/GaN HEMTs. 

In summary, we have demonstrated a 90 nm T gate $\mathrm{\beta} $-Ga\textsubscript{2}O\textsubscript{3} MOSFET  with process optimization to eliminate the regrowth interface resistance. We extracted near 55 GHz  f\textsubscript{MAX}, highest among $\mathrm{\beta} $-Ga\textsubscript{2}O\textsubscript{3} devices, with a gate-to-drain breakdown voltage of 125 V. However, degradation of f\textsubscript{T} and f\textsubscript{MAX} and current collapse  were observed in these un-passivated devices after large signal measurement trials which are possibly caused by trap introduction after repeated DC and RF measurements. Neverthles, the high f\textsubscript{MAX} is a significant achievement in terms of prospective applications of  
$\mathrm{\beta} $-Ga\textsubscript{2}O\textsubscript{3} in X band.

\section*{SUPPLEMENTARY MATERIAL}
See the supplementary material for breakdown analysis,  a detailed discussion on RF performance degradation before and after repeated DC, RF measurement. Scatter plot showing degradation of f\textsubscript{T}, f\textsubscript{MAX} after different measurements for 5 different devices, pulsed IV measurement for 200 ns pulse width for different bias points showing current collapse after stressing the device.

\begin{acknowledgments}
We acknowledge the support from AFOSR (Air Force Office of Scientific Research) under award FA9550-18-1-0479 (Program Manager: Ali Sayir), from NSF under awards  ECCS 2019749, 2231026 from Semiconductor Research Corporation under GRC Task ID 3007.001, and II-VI Foundation Block Gift Program. This work used the electron beam lithography system acquired through NSF MRI award ECCS 1919798. We would like to thank Dr. Andrew Green and Dr. Neil Moser from the Air Force Research Laboratories for their help in large signal measurements. 
\end{acknowledgments}

\section*{Data Availability Statement}

The data that support the findings of this study are available from the corresponding author upon reasonable request.


\section{REFERENCES}
\nocite{*}
\bibliography{aipsamp}

\end{document}